\documentclass[manuscript]{acmart}
\AtBeginDocument{%
  }

\setcopyright{acmlicensed}
\copyrightyear{2025}
\acmYear{2025}
\acmDOI{XXXXXXX.XXXXXXX}
\acmISBN{978-1-4503-XXXX-X/2018/06}

\begin{document}

%%
%% The "title" command has an optional parameter,
%% allowing the author to define a "short title" to be used in page headers.
\title{It’s Not Just Labeling: A Research on LLM Generated Feedback Interpretability and Image Labeling Sketch Features}

%%
%% The "author" command and its associated commands are used to define
%% the authors and their affiliations.
%% Of note is the shared affiliation of the first two authors, and the
%% "authornote" and "authornotemark" commands
%% used to denote shared contribution to the research.
\author{Baichuan Li}
\email{bl30@tamu.edu}
\orcid{0009-0009-1432-6523}
\author{Larry Powell}
\email{larry.powell@tamu.edu}
\orcid{0000-0001-9525-0405}
\author{Tracy Hammond}
\email{hammond@tamu.edu}
\orcid{0000-0001-7272-0507}
\affiliation{%
  \institution{Texas A\&M University}
  \city{College Station}
  \state{Texas}
  \country{USA}
}

%%
%% By default, the full list of authors will be used in the page
%% headers. Often, this list is too long, and will overlap
%% other information printed in the page headers. This command allows
%% the author to define a more concise list
%% of authors' names for this purpose.
\renewcommand{\shortauthors}{Baichuan et al.}

%%
%% The abstract is a short summary of the work to be presented in the
%% article.
\begin{abstract}
  The accuracy of training data directly impacts the performance of lots of applications powered by machine learning models, influencing their ability to make reliable predictions. In industries such as transportation, healthcare, and robotics, precise image labeling is essential for building robust AI systems. Domains like autonomous driving rely on well-annotated datasets to enhance object detection models, ensuring safer and more efficient navigation. Similarly, in medical applications like organ detection and disease diagnosis, accurate labeling is critical for training models to assist healthcare professionals. However, traditional labeling methods, such as manually drawing bounding boxes, are labor-intensive and require significant technical expertise. Furthermore, these methods and their associated annotation pipelines lack real-time feedback, increasing the risk of inconsistent or low-quality annotations. This limitation makes it difficult for non-experts and novices to contribute effectively, restricting the scalability and accessibility of high-quality data collection. This research proposed the concept of utilizing free-hand sketches as labeling approach and LLM as feedback provider to streamline the labeling process. This approach lowers both cognitive and technical barriers, enabling a broader range of users to participate in annotation tasks without extensive training. The study examines the relationship between sketch recognition features and LLM evaluation metrics using a synthetic dataset. By analyzing how different sketching styles affect LLM-generated feedback, this research seeks to improve the reliability and adaptability of LLM-assisted labeling. Understanding these relationships can help reduce uncertainty in LLM-based image annotation and improve the explainability of LLM-generated feedback through sketch input. The main contribution of this research is the development of a sketch-based virtual assistant powered by LLMs, designed to simplify the annotation process for non-expert users. Additionally, this study explores the impact of different prompting strategies on LLM performance, examining how variations in sketch features influence LLM-generated outputs. These findings will contribute to the development of more effective LLM-driven annotation tools, making image labeling more accessible, scalable, and explainable across multiple domains.
\end{abstract}

%%
%% The code below is generated by the tool at http://dl.acm.org/ccs.cfm.
%% Please copy and paste the code instead of the example below.
%%


%%
%% Keywords. The author(s) should pick words that accurately describe
%% the work being presented. Separate the keywords with commas.

% \received{20 February 2007}
% \received[revised]{12 March 2009}
% \received[accepted]{5 June 2009}

%%
%% This command processes the author and affiliation and title
%% information and builds the first part of the formatted document.
\maketitle

\section{Introduction}

\subsection{Motivation}
Accurate and precise image labeling is crucial to supervised machine learning models in several domains like autonomous vehicles, medical diagnosis systems, and intelligent manufacturing industries \cite{sager2021survey}. The need for high-quality annotations has grown as autonomous systems rely on vast datasets with accurately labeled objects to achieve reliable object detection and segmentation based on deep learning \cite{montserrat2017training}. Traditionally, image annotation has been accomplished through manual techniques such as drawing bounding boxes or polygonal outlines around objects by dragging and pointing. Moreover, not too many of the current labeling platforms provide feedback to users, so that they are not aware of the quality of the outcomes. Therefore, it is necessary to explore new methods that enable image annotators, especially non-experts, to contribute efficiently through interactive, feedback-assisted approaches.

Most existing solutions are based on mouse-oriented gestures like dragging and pointing. Image labeling platforms such as Roboflow have functions such as “smart polygon” tools, which can automatically draw a polygon according to the understanding of the AI assistant based on multiple points drawn by the user \cite{mahmud2024advancing}. However, labeling platforms such as this require a huge amount of learning costs and do not provide users with feedback.

Based on existing solutions, several critical challenges remain unresolved. Manual annotation methods continue to be labor-intensive and time-consuming \cite{hanbury2008survey}. Automated tools powered by artificial intelligence lack the flexibility to handle various real-world conditions, since they can sometimes miss fine structures and hallucinate small disconnected components \cite{kirillov2023segment}. Therefore, it is important to have an intuitive, user-friendly solution that reduces the cognitive and technical load on users, allowing them to produce consistent, high-quality annotations without extensive training. Therefore, this research aims to discover the possibility of using a sketch-based interface and LLM to enhance image labeling performance through a synthetic dataset. 

\subsection{Domain Context}
As we know, training a machine learning or deep learning model requires plenty of training data with high quality \cite{montserrat2017training}. The image labeling domain is integral to computer vision applications, enabling supervised machine learning models to recognize and interpret visual data by mapping spatial information in images with semantic labels. This process provides essential training data for computer vision systems used in areas like autonomous driving \cite{feng2020deep}, healthcare diagnostics \cite{elakkiya2021cervical}, and even micropalaeontology \cite{kirillov2023segment}. While traditional labeling methods are accurate, they are highly labor-intensive, requiring considerable human expertise and time \cite{hanbury2008survey}. Current efforts focus on developing interactive labeling tools that support both manual and semi-automated labeling, with enhancements in user interfaces, collaborative features, and gamification elements. Despite advancements, labeling remains a challenging and resource-intensive task, prompting continued research into efficient, scalable methods that can provide high-quality, large-scale annotations \cite{sager2021survey}. LabelMe \cite{russell2008labelme} is an iconic example. It is a platform where users can import images, segment objects by using polygons, add labels, and finally output it to XML files \cite{rapson2018reducing}, which can be used for model training in the future.

LLM has been proven to be an effective tool for assisting humans, as prompt-based modules can significantly reduce deployment time for complex tasks \cite{taneja2024can}. Using LLMs for annotation tasks is cost-effective, but compared to human annotation, they may introduce bias and errors, especially in domain-specific and complex tasks \cite{wang2024human}. Therefore, combining human expertise with LLMs is ideal to ensure precise and reliable annotations \cite{wang2024human}. Specifically, for LLM-assisted image labeling, the relationship between LLM outputs and labeling strokes requires further exploration.

\subsection{Summary of Solution}
To address the challenges of image labeling, this research focuses on exploring the feasibility of allowing users to label images using free-hand sketches while receiving feedback from an LLM. The proposed sketch-based assistant enables users to draw freely around objects using a stylus and tablet, leveraging the LLM to provide feedback based on labeled images and prompts. Figure \ref{ideal} below illustrates the ideal workflow of this system. The rest of this paper will elaborate on the process highlighted in the red box, detailing how images labeled through free-hand sketches influence LLM interpretability and how different prompts affect the evaluation feedback generated by the LLM by using a synthetic dataset.

\begin{figure}[h]
\centering
\includegraphics[width=\linewidth]{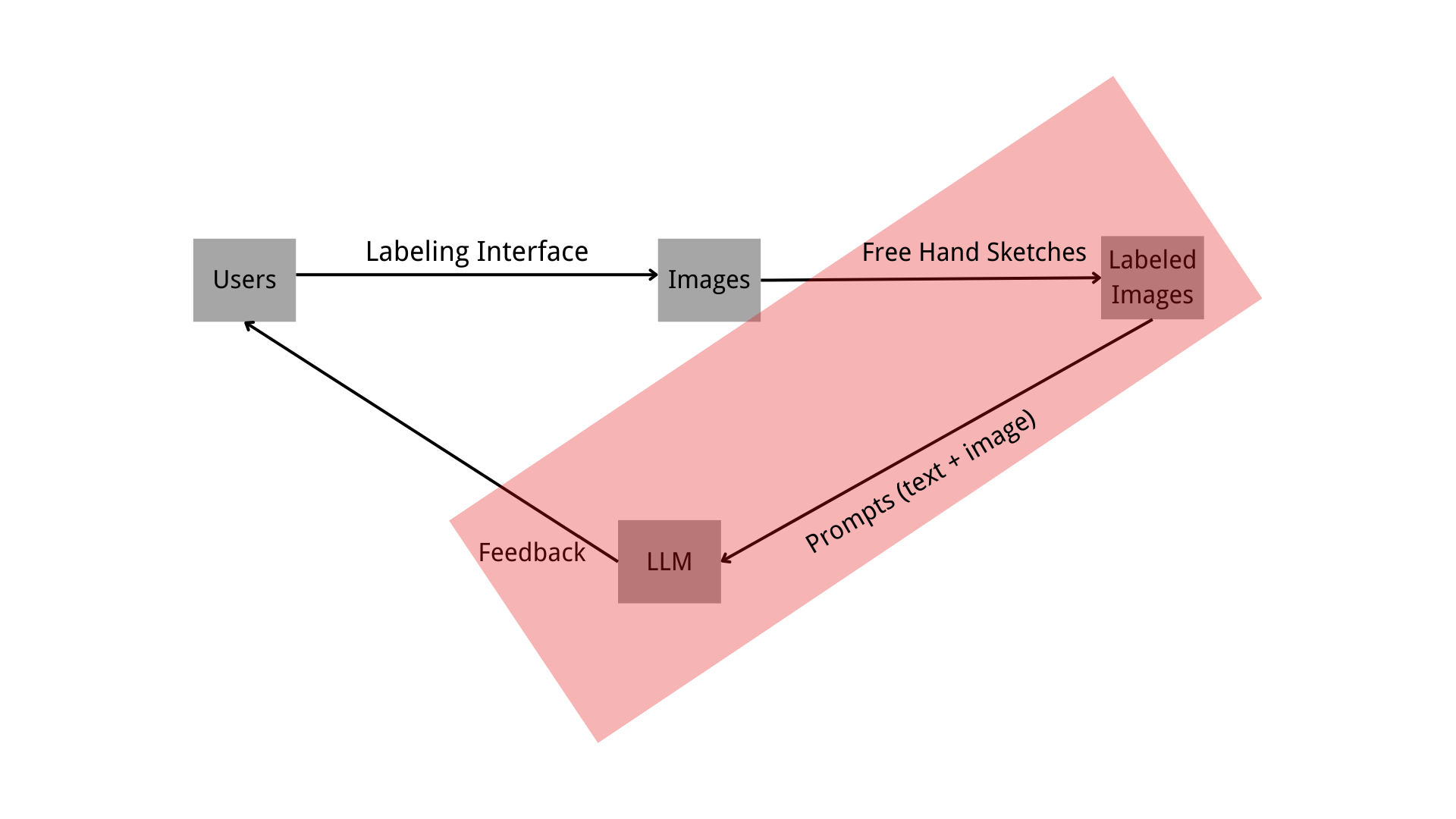}
\Description{Diagram showing the ideal flow of a sketch-based image labeling system.}
\caption{This diagram shows the ideal flow of sketch-based image labeling system. Users can utilize sketch-based image labeling interface to draw free-hand sketches and receive feedback from an LLM. This research mainly focused on the relationship between labeled images and LLMs, which is highlighted in the red box.  }
\label{ideal}
\centering
\end{figure}

\subsection{Research Questions}
This research focuses on 2 research questions: 

\begin{itemize}
\item \textbf{RQ1}: How much sketch features from synthetic dataset can influence the interpretability of LLM’s feedback generation by using different prompting strategies?
\item \textbf{RQ2}: What are the trade-offs among different prompting strategies for LLMs to evaluate images labeled by sketches. 
\end{itemize}

To define the success of this research, several key steps must be taken. First, it is essential to generate a synthetic dataset that represents stroke-based labeling as 2D points within an image dataset. This dataset should closely simulate human labeling behavior to ensure realistic testing conditions. Next, a valid method must be established to quantify both the characteristics of sketch-based labeling and the accuracy of LLM-generated feedback. This includes extracting the features of the sketches in conveying labeling intent, as well as assessing the reliability of LLM feedback in interpreting these sketches. Finally, the results should demonstrate the correlation between image labeling and LLM-generated feedback. Additionally, the study should identify significant differences in LLM evaluation metrics when feedback is generated using different prompting strategies, highlighting how various prompt designs influence interpretability of LLM.

\section{Related Work}
\subsection{Image Labeling Platform}
LabelMe \cite{russell2008labelme} is a web-based tool and database designed for image annotation, primarily used to build a large collection of images with labeled objects for computer vision research. It allows users to annotate images by drawing polygonal boundaries around objects, supporting detailed labeling with minimal supervision. LabelMe is a traditional tool for labeling images, and this research will focus on sketch-based labeling and provide feedback to users.
 
Panoptic Image Annotation \cite{uijlings2020panoptic} presents a collaborative annotation framework where a human annotator and an automated assistant work together to efficiently create panoptic segmentation labels. The assistant leverages a set of predefined segments generated by Mask R-CNN and uses contextual cues from the annotator’s actions to intelligently predict and modify other segments in the image. The study demonstrates that their method is faster without compromising label quality when tested on COCO and ADE20k datasets. 
 
Label Assistant \cite{schilling2021label} presents a sketch-based labeling tool to assist the annotation process in image segmentation tasks, especially for deep learning applications. The proposed workflow integrates various methods, including pre-labeling, image pre-processing, and post-assistance to aid annotators in labeling complex image datasets efficiently. The authors demonstrate the effectiveness of the workflow on two biomedical image segmentation datasets, showing that their approach reduces annotation effort while maintaining high-quality labels. However, they didn’t provide feedback to users. 
 
More Efficient Labeling Tool (MELT) \cite{rapson2018reducing} designed to streamline the creation of ground-truth labels for image segmentation tasks. MELT incorporates several novel features, such as automatic zoom to existing bounding boxes and the ability to track arbitrarily shaped objects across video frames. MELT enhances the efficiency of manual labeling by integrating features like tracking and automatic zoom, however, this approach has some false positive or false negative cases when two objects are overlapping, but sketch-based labeling can solve this issue.

The Segment Anything Model (SAM) \cite{kirillov2023segment} introduces a prompt-based segmentation model, capable of generating object masks from various prompts like points, boxes, and text. It is a very powerful model, and there are some existing labeling platform that uses this model to auto-label objects. Even though it can automatically separate objects from images, it relies on a large volume of training datasets, and its performance in real-time applications is limited by the heavy image encoder. Moreover, it is designed for general segmentation tasks, so it may underperform in highly specialized domains \cite{kirillov2023segment}. 

Raghavan et al. \cite{raghavan2021virtual} presents an Android application designed to assist individuals with partial vision impairment by leveraging AI for real-time object detection and a voice-based chatbot. The system works well under good lighting conditions. Therefore, integrating AI feedback into our system is available and necessary. 

\subsection{Prompt Engineering}

White et al. \cite{white2023prompt} mentioned the importance of prompts for LLMs, as they greatly influence subsequent interactions and outputs by providing specific rules and guidelines. Meanwhile, this paper provides a framework that documents patterns of structuring prompts to resolve different tasks. More importantly, prompt patterns can provide reusable solutions to specific problems. Basically, there are five prompt patterns: Input Semantics, Output Customization, Error Identification, Prompt Improvement, and Interaction. For the Persona Pattern in the Output Customization category, LLMs are expected to adopt a persona and hold a certain point of view or perspective. In the design of the prompt, the user should inform the LLM of their persona and provide the desired guidelines or examples of output. In our paper, the LLM will play a role in judging or assisting the user in generating better results. We used a similar strategy to design prompts, but our input will include labeled images.

Tian et al. \cite{tian2024examining} analyzed the difference and trade-off between different prompting strategies through qualitative analysis. Prompts can be separated into instructions which include a task description, rubric statement output expectations, examples, and inputs. Meanwhile, based on the components, the prompting strategy can be classified into four templates: Few-Shot Rubric, Zero-Shot Rubric, Few-Shot Basic, and Zero-Shot Basic. In our research, we utilized the four prompting strategies in our prompt design. 

Vujinović et al. \cite{vujinovic2024using} explored the effectiveness of Chat-GPT in replacing humans as dataset annotators. It leverages prompt engineering to instruct Chat-GPT on how to label student states with selected actions. The result of the study reaches 95\% correctness, which proved that LLMs are capable of handling annotation-related tasks. Meanwhile, this study developed a reusable list of pedagogical actions applicable to various tutoring systems. The prompts first outlined the role of the LLM, then provided it with the action list it needed to follow, and finally defined a set of explicit rules that the LLM must obey.

Karakaya \cite{karakaya2025human} concludes prompt engineering strategies in human-AI interaction with LLMs so that users can have more precise results, especially for difficult and complex tasks. Key strategies include providing clear instructions, specifying scope and length, decomposing complex tasks using techniques like Chain-of-Thought prompting for better reasoning transparency, tailoring prompts to the audience, and assigning personas to ensure domain-specific responses. 

\subsection{Synthetic Dataset for Gestures and Sketches}

Caramiaux et al. proposed Gesture Variation Follower \cite{caramiaux2014adaptive}, a synthetic gesture dataset and an adaptive recognition system designed for interactive applications involving gesture variations. Their approach involves template-based recognition coupled with continuous, real-time adaptation to gesture variations such as speed, amplitude, and rotation. By employing Sequential Monte Carlo inference, their method continuously updated the estimated parameters and recognition results, allowing real-time tracking and adaptation of gestures. Evaluations on synthetic datasets demonstrated that their system effectively adapted to variations, such as changes in rotation, speed, and scaling, outperforming traditional offline template-based methods by dynamically tracking gesture features during execution. 

Ashbrook and Starner introduced MAGIC \cite{ashbrook2010magic}, a gesture design tool for creating and testing motion gestures in interactive systems in an intuitive manner. The tool enabled designers to ensure gestures are not mistakenly triggered by common everyday movements, significantly improving the reliability and usability of gesture-based interactions. MAGIC employed a synthetic dataset called the Everyday Gesture Library (EGL), which contains gesture samples collected from real-world daily activities, enabling rapid testing and iterative refinement of gesture-based interfaces.

Taranta et al. introduced a synthetic data generation technique known as Gesture Path Stochastic Resampling (GPSR) \cite{taranta2016rapid}, designed for rapid prototyping of gesture recognition systems. GPSR generates synthetic 2D gesture data by applying non-uniform resampling to original gesture paths, creating realistic variations through lengthening and shortening sub-paths. This method is efficient, easy to implement, and significantly improves recognition accuracy across various datasets. The evaluation demonstrated that GPSR notably reduced recognition errors compared to other state-of-the-art synthetic data generation methods.

\subsection{LLM for Annotation Tasks}

Wang et al. proposed a collaborative annotation framework by using LLMs and human annotators to improve annotation accuracy and efficiency. \cite{wang2024human} In this research, LLMs first generated labels and provided rationalizations for their predictions. A separate verifier then assessed the quality of these annotations, and then selected potentially incorrect labels for human re-annotation. This human-LLM collaboration effectively balanced annotation cost and accuracy. Empirical studies demonstrated that combining human annotators with LLM-generated explanations significantly improved labeling performance, especially by identifying errors and uncertainties in LLM outputs, thus highlighting the importance of explanation quality and prompt strategies in collaborative annotation tasks.

Taneja and Goel proposed an active label correction method ALC3 \cite{taneja2024can} to improve annotation quality in datasets labeled by LLMs. Recognizing that LLM-generated annotations can be noisy, their approach combined auto-correction, human annotation, and filtering in an iterative approach. By selectively correcting examples identified as most likely mistakenly annotated, this method efficiently used human feedback and significantly reduced the amount of manual annotation required. Experiments across various tasks demonstrated that ALC3 achieved good performance comparable to fully-human-annotated datasets while requiring substantially less human intervention, underscoring its effectiveness for enhancing LLM-based modular AI systems.

\subsection{Evaluation of LLM-Generated Content}

Shahul et al. introduced RAGAS \cite{es2024ragas}, an automated evaluation framework specifically designed for assessing Retrieval Augmented systems, which combined retrieval modules with LLMs. RAGAS evaluated three key dimensions—faithfulness, answer relevance, and context relevance. By employing LLM-based evaluation metrics, RAGAS efficiently measured how accurately LLM-generated answers reflect retrieved context, how well answers addressed user queries, and how focused the retrieved context is. This approach significantly accelerated evaluation cycles, improving the development and deployment of robust RAG-based systems.

Van Schaik and Pugh discussed automated evaluation methods for summaries generated by LLMs \cite{van2024field}. They mentioned that traditional NLP metrics like ROUGE and BLEU remain useful but often fail to capture deeper semantic qualities. More recent methods, including LLM-based evaluators, have emerged to address this gap by evaluating qualities such as factual consistency, coherence, fluency, and relevance. However, LLM evaluators themselves faced challenges, such as positional bias, verbosity bias, and limited reasoning capabilities. They recommended combining multiple evaluation metrics, including both LLM-based and traditional methods, to robustly assess LLM outputs while also involving domain experts to validate the effectiveness and reliability of these metrics.

Awasthi et al. proposed HumanELY \cite{awasthi2023humanely}, a structured framework and web-based tool for the comprehensive human evaluation of LLM outputs. Because of the limitations of automated metrics, HumanELY focused on five evaluation dimensions: relevance, coverage, coherence, harm, and comparison, each assessed through detailed Likert-scale questions. Their method provided clear, consistent, and measurable evaluation criteria that address typical shortcomings such as bias, toxicity, hallucination, and lack of coherence in LLM-generated text. HumanELY emphasizes human judgment is crucial for assessing complex, context-dependent qualities of LLM outputs, particularly in critical domains such as healthcare. Compared to recent LLM-evaluation methods in the healthcare domain, HumanELY achieved the highest accuracy and lowest harm in a comprehensive manner.

\section{Methodologies}

This section describes the methods used in the research to generate synthetic sketches for image labeling, prompting strategies, and quantification of image labeling and LLM performance. The following Figure \ref{meths} is a brief overview of the workflow of the methodologies. 

\begin{figure}[ht]
\includegraphics[width=15cm]{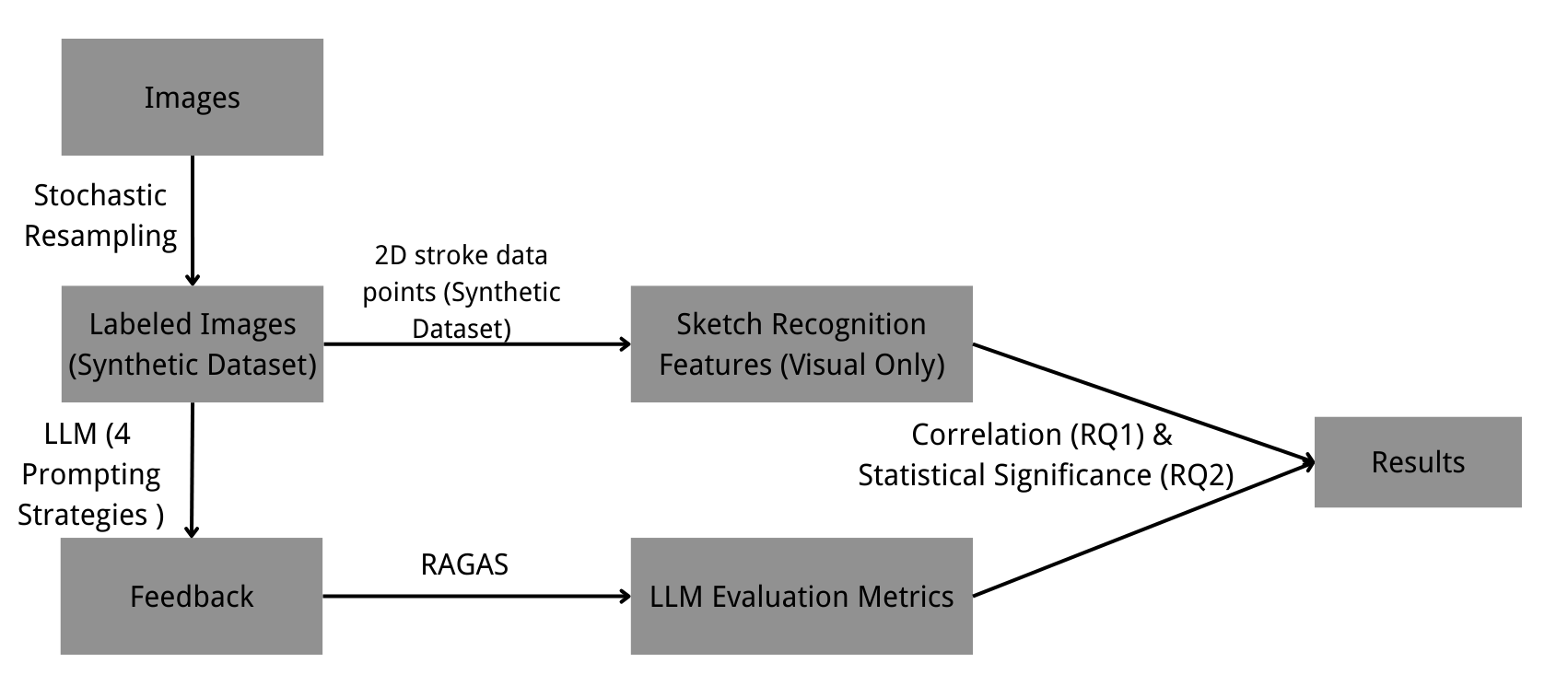}
\centering
\caption{This diagram is the methodology overview in this research. We firstly generated the synthetic dataset for labeled images through stochastic resampling, and then generate feedback by inputting four different prompts to the LLM. After that, SR features and LLM evaluation metrics are computed to achieve the quantification goal. Finally, the results were derived by statistical analysis.}
\centering
\label{meths}
\end{figure}

Initially, an image collection was selected and transformed into labeled images with corresponding stroke data through stochastic resampling. These stroke data points were represented as 2D coordinates, effectively simulating free-hand sketches around target objects within the images. This approach generated a synthetic dataset that realistically mimics human annotation behaviors, capturing the essential variability and complexity present in free-hand sketching.

Subsequently, the labeled images from the synthetic dataset were processed by leveraging an LLM, and several distinct prompting strategies were employed to elicit LLM-generated feedback for each labeled image. This feedback was intended to assess, and guide annotations, thus enhancing annotation precision and reliability. The variations in the prompt strategies are specifically designed to explore how different prompts influence the quality and relevance of feedback provided by the LLM.

Following this, data quantification steps were conducted. The synthetic stroke data points undergone feature extraction, focusing exclusively on visual attributes to derive sketch recognition features. These features captured crucial characteristics such as shape, curvature, and length, which may influence the LLM's interpretation of the labeled sketches. Then, feedback from the LLM was systematically evaluated by using an established LLM evaluation framework, which employs metrics such as relevance, faithfulness, and context alignment in the LLM outputs.

Finally, the extracted sketch recognition features and LLM evaluation metrics were statistically analyzed. The goals of this research were to examine whether certain sketch characteristics can influence LLM interpretability and explore how prompting strategies affect the nature and usefulness of LLM-generated feedback.

\subsection{Dataset}

The images used in this research included 110 selected images sourced from the Supervisely dataset \cite{superviselyDataset}. Each of these images contained one or more human objects, accompanied by corresponding masks that precisely delineate the human shapes from the background. These masks enabled accurate segmentation for subsequent sketch-based labeling generation. The Figure \ref{ex} and Figure \ref{ex_mask} below illustrate an example of a selected image and its associated mask.

\begin{figure}[h]
\centering
\includegraphics[width=\linewidth]{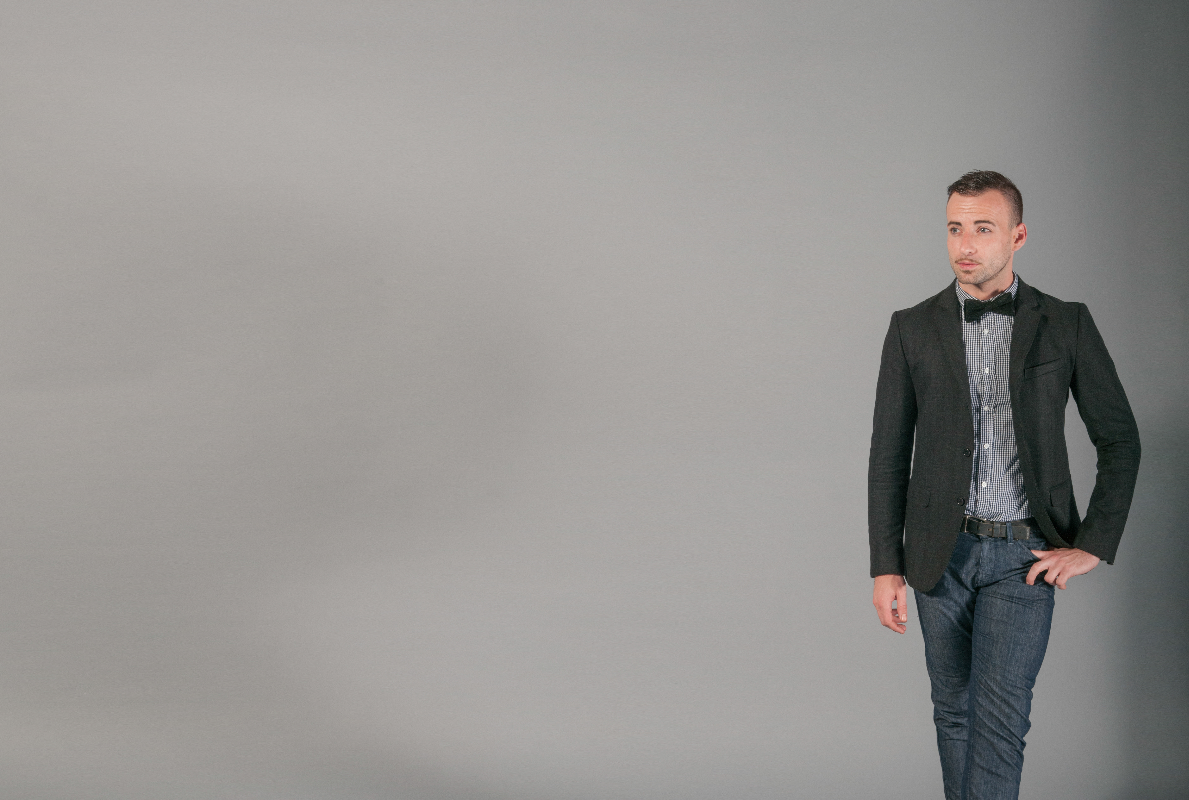}
\caption{This figure is an example image used for synthetic dataset generation. }
\centering
\label{ex}
\end{figure}

\begin{figure}[ht]
\centering
\includegraphics[width=\linewidth]{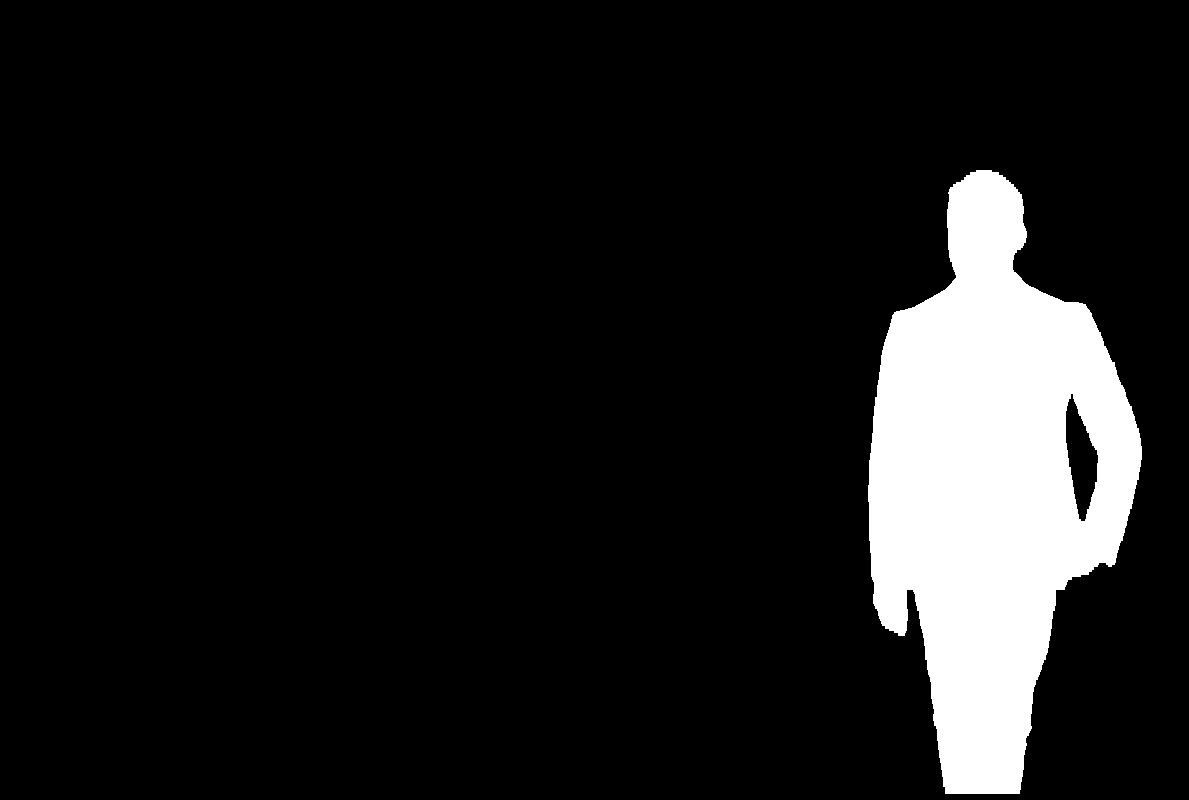}
\caption{This figure is an example image mask used for synthetic dataset generation. }
\centering
\label{ex_mask}
\end{figure}

\subsubsection{Synthetic Dataset for Sketch-Based Labeling Strokes}

To obtain images labeled by free-hand sketches, we used a synthetic dataset composed of lists of 2D points representing strokes typically drawn by users using a stylus and tablet. Several reasons justified the choice of a synthetic dataset in this research. Firstly, synthetic data can be generated rapidly in large quantities \cite{taranta2016rapid}, significantly enhancing efficiency and providing ample data for analysis. Additionally, collecting real user sketch data can be time-consuming and resource-intensive \cite{leiva2015gestures}; thus, as a pilot research employing synthetic data saves considerable time while still yielding valuable insights. Most importantly, as real user's sketch data can be highly varied, using synthetic dataset can provide controlled and quantitative variations \cite{caramiaux2014adaptive}. Lastly, using synthetic datasets effectively mitigates potential privacy and security issues \cite{taranta2016rapid}.

\begin{figure}[h]
\centering
\includegraphics[width=\linewidth]{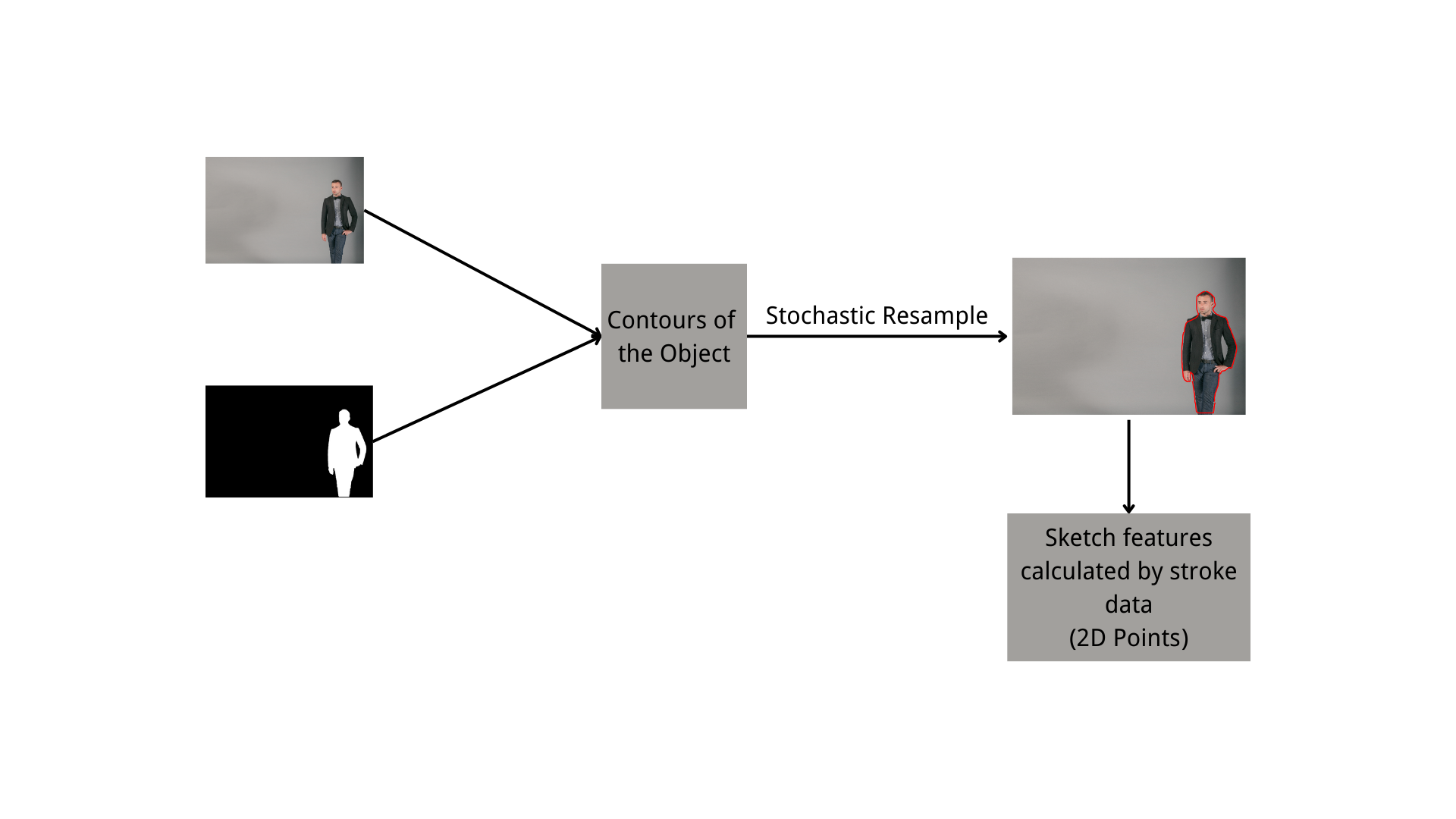}
\caption{This diagram shows the steps to generate synthetic dataset for labeling strokes. Firstly contours of objects of interest were extracted and performed stochastic resampling. Finally SR features were computed.}
\centering
\label{steps}
\end{figure}

The Figure \ref{steps} is a diagram that demonstrates the steps we can obtain the sketch-based labeling strokes and extract relevant features. 

The process began with two inputs: an image containing the object of interest and its corresponding mask, which precisely identified the object's boundary. These inputs were used to extract the object's contours. Next, the stochastic resampling technique was applied to generate free-hand sketches that simulate human-like sketching in image labeling based on the contours. Finally, the resampled contours were divided into multiple strokes represented as lists of 2D points, from which sketch features can be calculated in the next step.

\subsubsection{Stochastic Resampling}
The approach used in this research to generate the synthetic sketch stroke data to represent image labeling is gesture path stochastic resampling (GPSR) \cite{taranta2016rapid}. GPSR is a synthetic data generation technique for rapid prototyping and improving 2D gesture recognition systems. It offers several advantages, including computational efficiency, ease of implementation, and minimal coding overhead. It generates synthetic gesture samples by non-uniformly resampling original gestures, effectively creating realistic variations through subtle lengthening and shortening of gesture sub-paths. Meanwhile, evaluations verified that GPSR outperformed other state-of-the-art synthetic data generation methods in reducing recognition errors. Its simplicity, speed, and effectiveness made it particularly suitable for synthetic stroke data generation for image labeling in our research. 

Before performing stochastic resampling, the contours of objects of interest which each initially represented as lists of points were combined into a single flat list. This representation allowed subsequent operations to be applied uniformly across all strokes without concern for their original segmentation.

After flattening, the number of points for resampling (n) was determined either by a predefined constant resample\_cnt or by calling an optimization function optimal\_n that calculated the ideal number of points based on the specific dataset. Additionally, a small number of randomly chosen points remove\_cnt are subsequently removed from the resampled set to simulate realistic omissions or minor inaccuracies typically observed in free-hand sketches. Another constant used was variance, which determines the randomness of distances between points. In our research, we determined the number of points using the optimal n method, as it can generate more realistic and precise gesture data compared to using a static number \cite{taranta2016rapid}. The variance in this research was set to 0.25, which is the default value used in the original study \cite{taranta2016rapid}. All constants used in the stochastic resampling process are listed in Table \ref{tab:constants}.

% \begin{table}[h!]
% \centering
% \begin{tabular}{|p{3cm}|p{6cm}|p{2cm}|}
% \hline
% \textbf{Constant Name} & \textbf{Description} & \textbf{Value}\\ \hline
% use\_optimal\_n & Determines if optimal number of points is calculated dynamically & True \\ \hline
% resample\_cnt & Default number of points if optimal\_n is False & N/A \\ \hline
% remove\_cnt & Number of points randomly removed from resampled data & N/A \\ \hline
% variance & Randomness of distances between points in stochastic resampling & 0.25 \\ \hline
% \end{tabular}
% \caption{This table lists all the constants used in the stochastic resampling process.}
% \label{tab:constants}
% \end{table}

\begin{table}[h!]
\centering
\begin{tabular}{p{0.2\linewidth} | p{0.6\linewidth} | p{0.1\linewidth}}
\hline
\textbf{Constant Name} & \textbf{Description} & \textbf{Value}\\ \hline
use\_optimal\_n & Determines if optimal number of points is calculated dynamically & True \\ \hline
resample\_cnt & Default number of points if optimal\_n is False & N/A \\ \hline
remove\_cnt & Number of points randomly removed from resampled data & N/A \\ \hline
variance & Randomness of distances between points in stochastic resampling & 0.25 \\ \hline
\end{tabular}
\caption{This table lists all the constants used in the stochastic resampling process.}
\label{tab:constants}
\end{table}

The stochastic resampling procedure began by generating random intervals to create variability. Specifically, intervals between points were randomly generated based on a pre-defined constant variance, which determined the degree of randomness. These intervals were scaled so that their sum equals the total length of the original path, thus preserving overall gesture proportions while introducing local variability.

The points were then resampled based on these stochastic intervals. Iteratively, new points were calculated along the original gesture path, ensuring each new point was placed at distances corresponding to these randomized intervals. If the cumulative length exceeded an interval, linear interpolation was used to precisely determine the location of the new point. This continued until the desired number of resampled points was achieved, effectively generating a synthetic gesture with subtle variations.

Subsequently, the newly generated points undergo normalization. This step involved first computing the bounding box around the original points and calculating distances between consecutive resampled points. Points were then normalized by adjusting the distances between them to maintain a consistent spatial scale relative to the original bounding box, preserving gesture proportions and overall size consistency.

\begin{figure}[h]
\centering
\includegraphics[width=\linewidth]{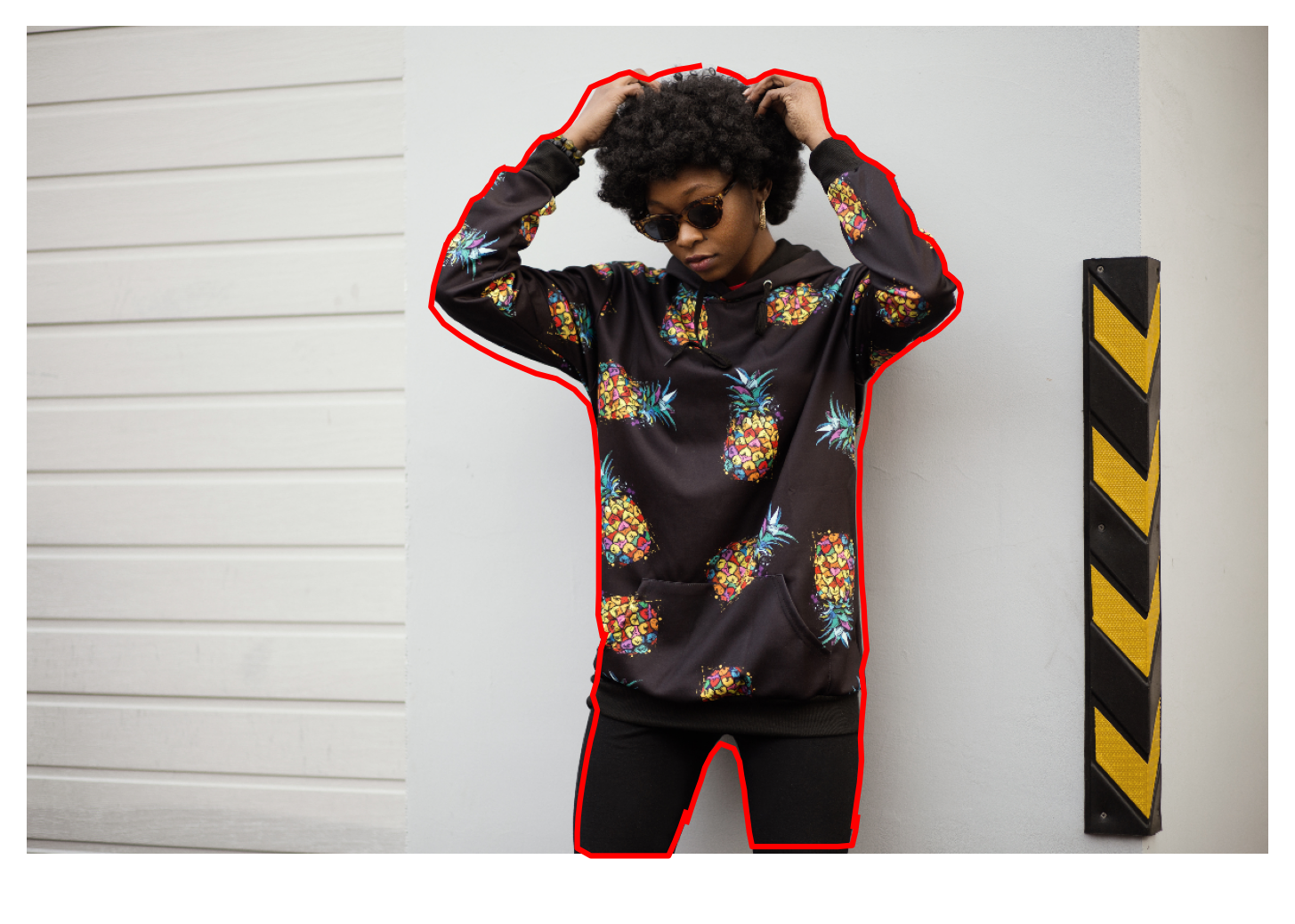}
\caption{This figure is an example labeled image from synthetic dataset generated by stochastic resampling.}
\label{sk1}
\centering
\end{figure}

\begin{figure}[h]
\centering
\includegraphics[width=\linewidth]{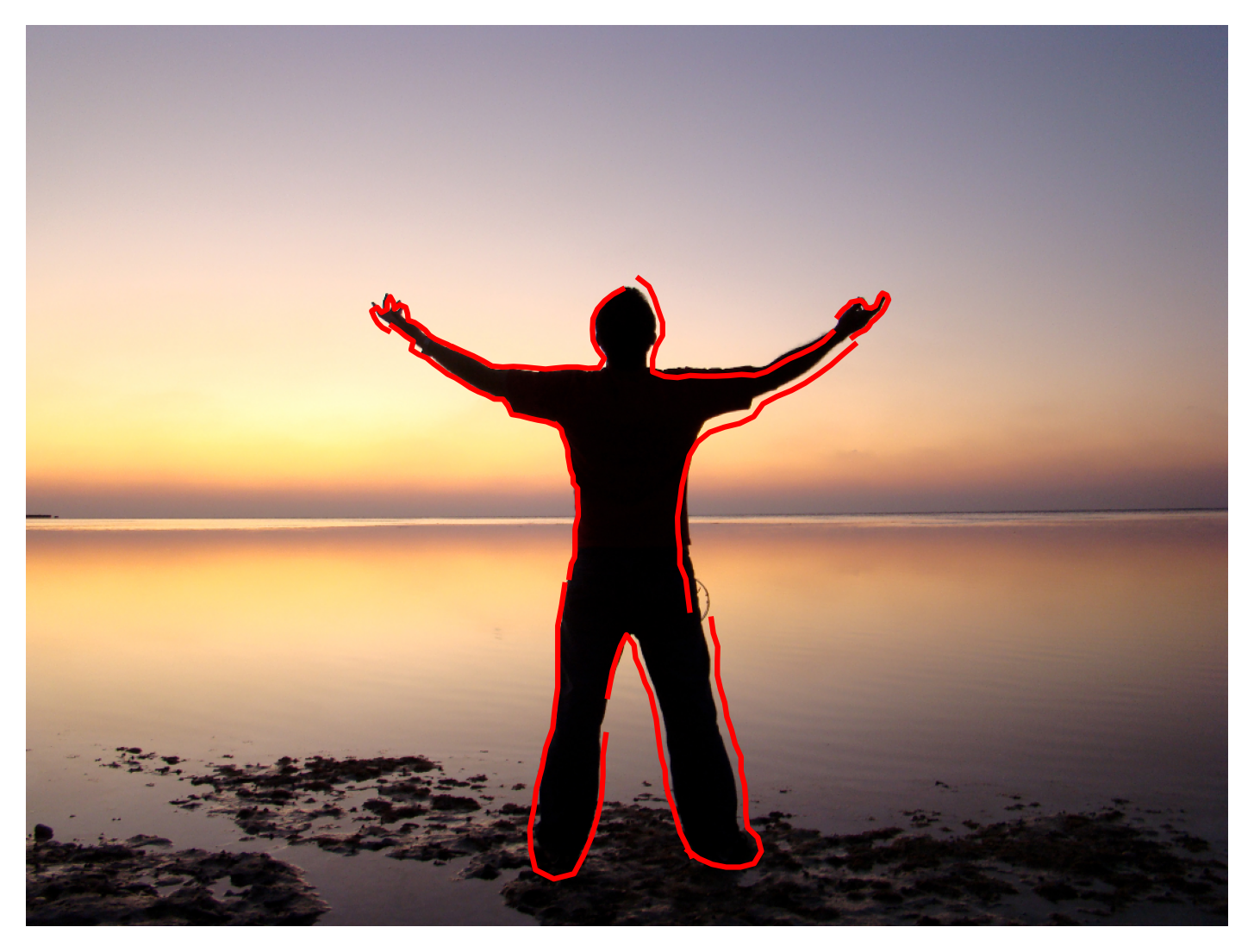}
\caption{This figure is an example labeled image from synthetic dataset generated by stochastic resampling.}
\label{sk2}
\centering
\end{figure}

Finally, the resampled and normalized points were repositioned so their center matched that of the original points. The repositioning was done by calculating the offset between the center of the original bounding box and the new normalized points, then shifting all points accordingly. The resulting points thus formed a synthetic stroke. To simulate human behavior as closely as possible, all stroke data points were sorted from left to right and top to bottom to reflect the drawing tendencies of right-handed users \cite{vlachos2004left}. Examples of generated sketch-based labeled images are shown in Figure \ref{sk1} and Figure \ref{sk2}.

\subsubsection{Sketch Recognition Features}

When sketch-based labeled images and their corresponding stroke data points are obtained, it is important to get the features that can mostly reflect the characteristics and properties of labeling sketches. In this research, Sketch Recognition (SR) features were calculated for each labeled object in each image. The SR features selected are shown in Table \ref{tab:sketch_features}, most of them were selected from the studies from Rubine \cite{rubine1991specifying} and Long (which is an extension of Rubine's work) \cite{long2000visual}, which primarily used to describe any single-stroke shape \cite{paulson2008paleosketch}.

\begin{table}[ht]
\centering
\begin{tabular}{p{0.35\linewidth} | p{0.6\linewidth}}
\hline
\textbf{Feature} & \textbf{Description} \\ \hline
R1 & Cosine of initial angle \\ \hline
R2 & Sine of initial angle \\ \hline
R3 & Length of bounding box diagonal \\ \hline
R4 & Angle of bounding box diagonal \\ \hline
R5 & Distance between first and last points \\ \hline
R6 & Cosine of angle between first and last points \\ \hline
R7 & Sine of angle between first and last points \\ \hline
R8 & Total gesture length \\ \hline
R9 & Total angle traversed \\ \hline
R10 & Absolute sum of angles traversed \\ \hline
R11 & Squared sum of angles traversed \\ \hline
L14 & Total angle traversed / stroke length (Average Rotation) \\ \hline
L15 & Stroke length / distance between endpoints (Density) \\ \hline
L16 & Stroke length / bounding box (second density) \\ \hline
L17 & Distance between endpoints / bounding box diagonal length (non subjective openness) \\ \hline
L20 & Total angle / absolute total angle (change in rotational motion) \\ \hline
Distance between strokes & Openness of the whole labeling \\ \hline
\end{tabular}
\caption{ This table lists all the sketch recognition features computed for each labeled image. The first eleven features were selected from Rubine \cite{rubine1991specifying}. The following five features were from Long \cite{long2000visual}. The last one measures the distance between each stroke in one piece of labeling, reflecting the openness of the whole labeling.}
\label{tab:sketch_features}
\end{table}

There are several reasons we used SR features to describe sketch-based labeling. First, it enables the quantification of labeling gestures by capturing key visual attributes such as stroke length, curvature, and stroke count, allowing numerical comparisons of gestures across different conditions. Second, sketch features facilitate understanding the difficulty and complexity associated with gesture articulation by measuring attributes like the number of strokes \cite{rubine1991specifying}, particularly beneficial for evaluating performance differences between user groups, such as those with and without motor impairments \cite{rekik2014understanding}. Finally, sketch-based analysis effectively reflects variations among stroke gestures, clearly distinguishing performance differences, thus supporting more informed design and evaluation of accessible gesture-based interactions \cite{vatavu2022understanding}. 

\subsection{Feedback Generation}

The feedback generation process for each labeled image involves interacting with OpenAI's GPT-4 API by carefully structuring the input prompts. We used four different prompting strategies from the study from Tian et al \cite{tian2024examining}. Basically, each prompt included both text components and labeled images. Leveraging this comprehensive set of inputs, for each image, LLM generated four pieces of feedback to evaluate or correct the labeled image annotations through each prompting strategy. 

\subsubsection{Prompt Design}
The four prompting strategies used in this research were based on the study from Tian et al \cite{tian2024examining}. As shown in Table \ref{comp}, for each prompt, there were basically four essential components. Task description and output expectations were required, while rubric statements and examples were optional. The task description explicitly outlines the evaluation goal and clearly communicates the task to the model. The rubric statement specified the detailed evaluation criteria structured around a defined scale. The output expectations guided how the model should format and structure its response to align with predefined standards. And finally, examples that provided explicit cases or demonstrations to help the model better understand the desired output format. All the prompts and referenced labeled images can be found in Appendix.

\begin{table}[!ht]
\centering
\begin{tabular}{p{0.35\linewidth} | p{0.6\linewidth}}
\hline
\textbf{Prompting Strategy} & \textbf{Components} \\ \hline
Zero-shot Basic & Task Descriptions, Output Expectations, Input \\ \hline
Zero-shot Rubric & Task Descriptions, Output Expectations, Rubric, Input \\ \hline
Few-shot Basic & Task Descriptions, Output Expectations, Example, Input \\ \hline
Few-shot Rubric & Task Descriptions, Output Expectations, Rubric, Example, Input \\ \hline

\end{tabular}
\caption{This table lists all four prompting strategies and their components. The task descriptions and output expections are required, but the rubric and example are optional. }
\label{comp}
\end{table}

Based on the selection of prompt components, there are four distinct prompting strategies: Zero-shot Basic, Zero-shot Rubric, Few-shot Basic, and Few-shot Rubric \cite{tian2024examining}. Zero-shot Basic provides only the fundamental instructions for the task along with output expectations and input, without any explicit rubric or examples. Zero-shot Rubric extends the basic strategy by explicitly providing a rubric statement alongside the task instructions and output expectations but still excludes examples. Few-shot Basic supplements the basic strategy by including specific examples that demonstrate the desired outputs, although it doesn't include a rubric statement. Finally, Few-shot Rubric represents the most comprehensive approach by incorporating all four components—task description, rubric statement, output expectations, explicit examples, as well as input—to clearly guide the model's outputs. Specifically, the rubrics were designed across four dimensions: whether the labeling completely encloses the object, whether the sketch is outside the image boundary, whether the sketch overlays the object of interest and the gap between the labeling sketches and the object. And the example used is a perfectly labeled image in which sketches are directly derived from the object's contours.

\subsubsection{LLM Evaluation Metrics}

After generating feedback for each labeled image by using each prompting strategy, it is necessary to evaluate and quantify the accuracy, relevance, and precision of the content generated. The pipeline used is RAGAS \cite{es2024ragas}. The input of RAGAS consisted of four components: the question posed by the user, contextual information, the answer generated by the LLM, and ground truth. The ground truth was derived by manually evaluating each labeled image based on the openness, boundary accuracy, alignment, and overlay criteria, following the rubric or expectations outlined in the prompts. The output from RAGAS includes three metrics \cite{es2024ragas}. First, context precision ensured that the generated answer contained only relevant and necessary information, whereas a higher score indicates fewer irrelevant details. Second, faithfulness measured whether the claims made in the generated feedback can be accurately inferred from the provided context, with a higher value indicating greater factual accuracy. Lastly, answer relevancy assessed how directly the generated feedback addresses the user's question, where a higher score means the answer is more closely aligned with the initial inquiry. These metrics together provided a comprehensive evaluation of the quality and effectiveness of LLM-generated feedback.

\section{Results}
\subsection{Data Analysis}
After calculating SR features and evaluation metrics based on the four prompting strategies for each image in the original image collection, we analyzed the dataset to answer the research questions. The first research question examined the correlation between SR features and the interpretability of LLM feedback, while the second investigated the trade-offs among different prompting strategies for evaluating sketch-based image labeling.

% \subsection{Correlation Exploration}

\subsubsection{Distance Correlation}

\begin{figure}[h]
\centering
\includegraphics[width=\linewidth]{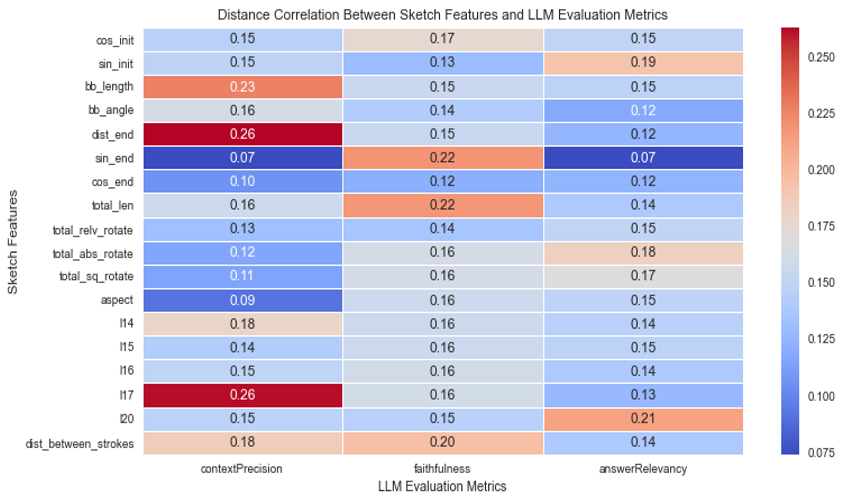}
\caption{This figure is the distance correlation results for zero-shot rubric, overall no strong correlation between SR features from the synthetic dataset and LLM evaluation metrics.}
\label{dist1}
\centering
\end{figure}

\begin{figure}[H]
\includegraphics[width=15cm]{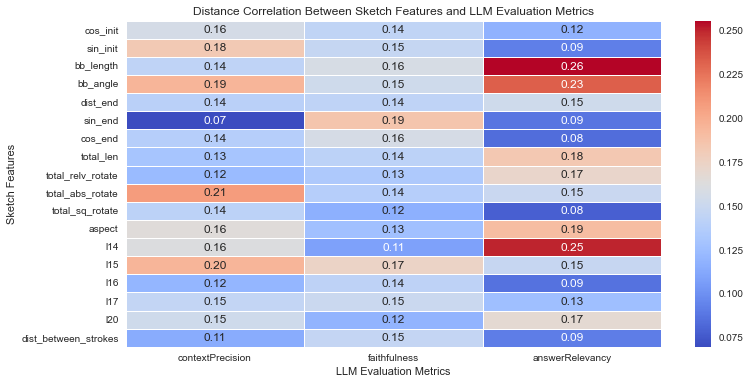}
\centering
\caption{This figure is the distance correlation results for zero-shot basic, overall no strong correlation between SR features from the synthetic dataset and LLM evaluation metrics.}
\label{dist2}
\centering
\end{figure}

\begin{figure}[h]
\centering
\includegraphics[width=\linewidth]{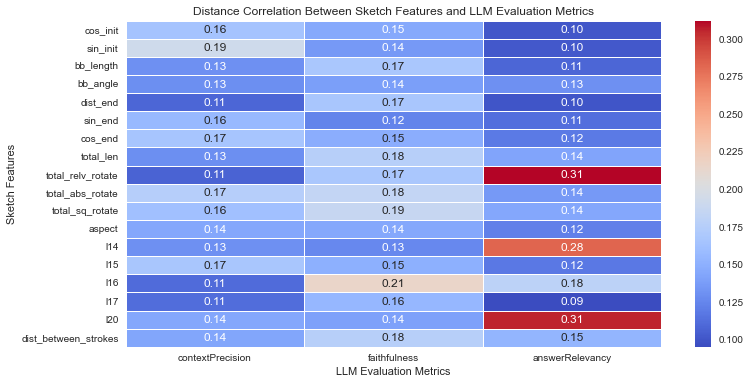}
\caption{This figure is the distance correlation results for few-shot rubric, overall no strong correlation between SR features from the synthetic dataset and LLM evaluation metrics.}
\label{dist3}
\centering
\end{figure}

\begin{figure}[h]
\centering
\includegraphics[width=\linewidth]{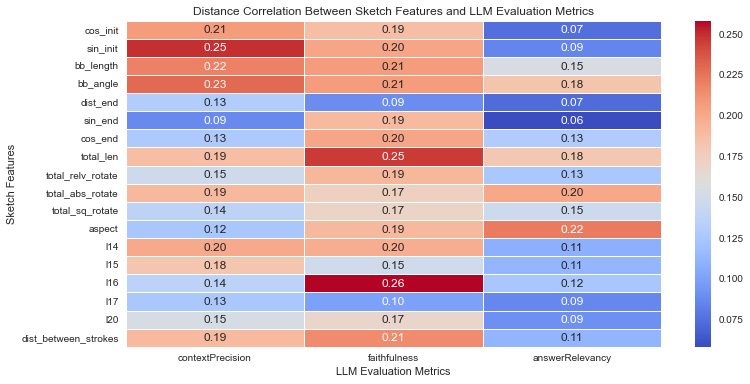}
\caption{This figure is the distance correlation results for few-shot basic, overall no strong correlation between SR features from the synthetic dataset and LLM evaluation metrics.}
\label{dist4}
\centering
\end{figure}

To answer the first research question, we first calculated the distance correlation between SR features and each LLM-evaluation metric generated by RAGAS for all prompting strategies. The results is shown in Figure \ref{dist1}, \ref{dist2}, \ref{dist3}, \ref{dist4}. Generally, specific sketch features can weakly predict the interpretability metrics of the LLM feedback.

For zero-shot rubric strategy, the distance correlation analysis in Figure \ref{dist1} primarily showed weak correlations overall. Specifically, the context precision metric exhibited moderately weak correlations with certain SR features, notably the distance between endpoints, L17 (the ratio of distance between endpoints to bounding box diagonal length), and the bounding box diagonal length itself. Faithfulness also demonstrated moderately weak correlations with several features, including the sine of the angle between endpoints, the total gesture length, and the distance between strokes. Additionally, the answer relevancy showed a weak correlation with feature L20, representing the ratio of total angle to absolute total angle. 

For zero-shot basic strategy, the distance correlation analysis in Figure \ref{dist2} revealed bounding box length (bb\_length) and L14 (average rotation) had a weak correlation with answer relevancy, suggesting that bounding box dimensions may slightly influence how well LLM feedback aligns with the intended question. Other sketch features displayed negligible correlations across different evaluation metrics. 

For few-shot rubric strategy, the distance correlation analysis in Figure \ref{dist3} demonstrated answer relevancy has a weak correlation with total relative rotation and total angle to absolute angle ratio (L20), suggesting that rotational complexity can somehow impact how well the LLM aligns its feedback with the input. While faithfulness and context precision exhibited generally negligible correlations across most features. 

Finally, for few few-shot basic strategy, distance correlation analysis in Figure \ref{dist4} demonstrated context precision has a weak correlation with sin\_init (sine of initial angle), suggesting that the initial stroke angle can influence how relevant the generated feedback is. Faithfulness exhibited its weak correlation with L16 (second density) and total stroke length, indicating that the size and density of the labeling sketch may slightly impact the factual accuracy of LLM-generated responses. Otherwise, the rest correlations were negligible.  

% \subsubsection{Pearson Correlation}

\subsubsection{Statistical Tests}
The second research question explored the trade-offs between different prompting strategies in evaluating sketch-based image labeling. We performed statistical tests to gain insights. At first, for each prompting strategy, we examined the distribution and normality of each LLM evaluation metric by drawing the histograms and performing the Shapiro–Wilk test.

\begin{figure}[h]
\centering
\includegraphics[width=\linewidth]{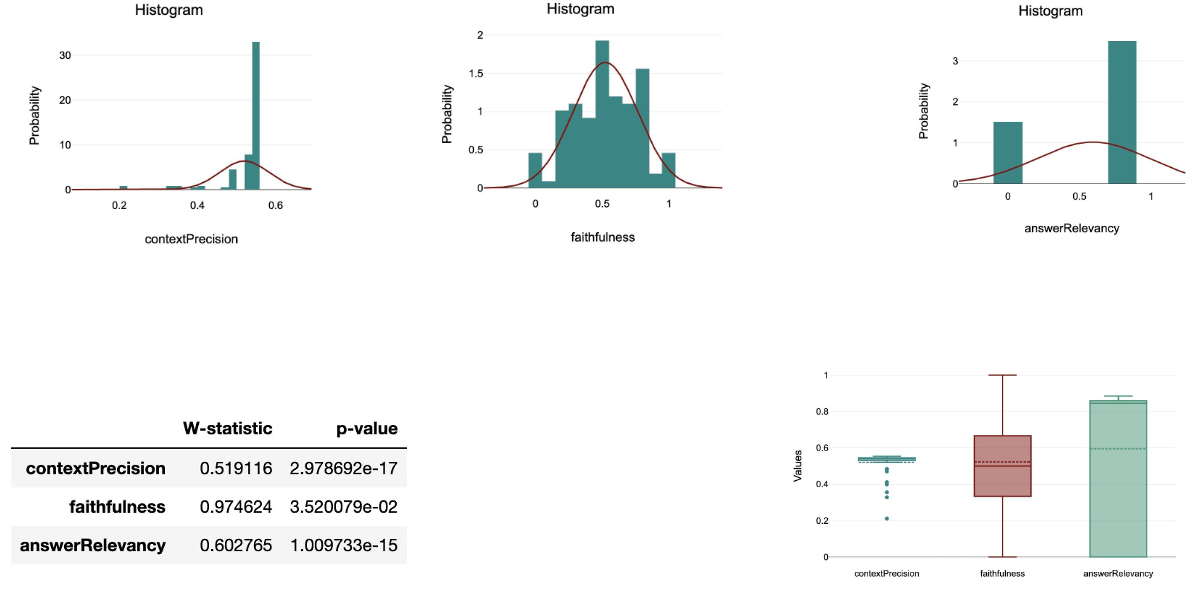}
\caption{This figure demonstrates the normality check for zero-shot basic, all the LLM-evaluation metrics are not normally distributed.}
\label{nom1}
\centering
\end{figure}

\paragraph{Normality Check} 
Looking at the LLM evaluation metrics for zero-shot basic in Figure \ref{nom1}, context precision histogram was highly skewed and indicating a dominant cluster and a long left tail. The faithfulness distribution is more balanced and slightly skewed. In contrast, the answer relevancy histogram exhibited a bimodal pattern. For the Shapiro–Wilk test, all the p-values of LLM evaluation tests are less than the significance level (0.05), which means they are not normally distributed. 

\begin{figure}[h]
\centering
\includegraphics[width=\linewidth]{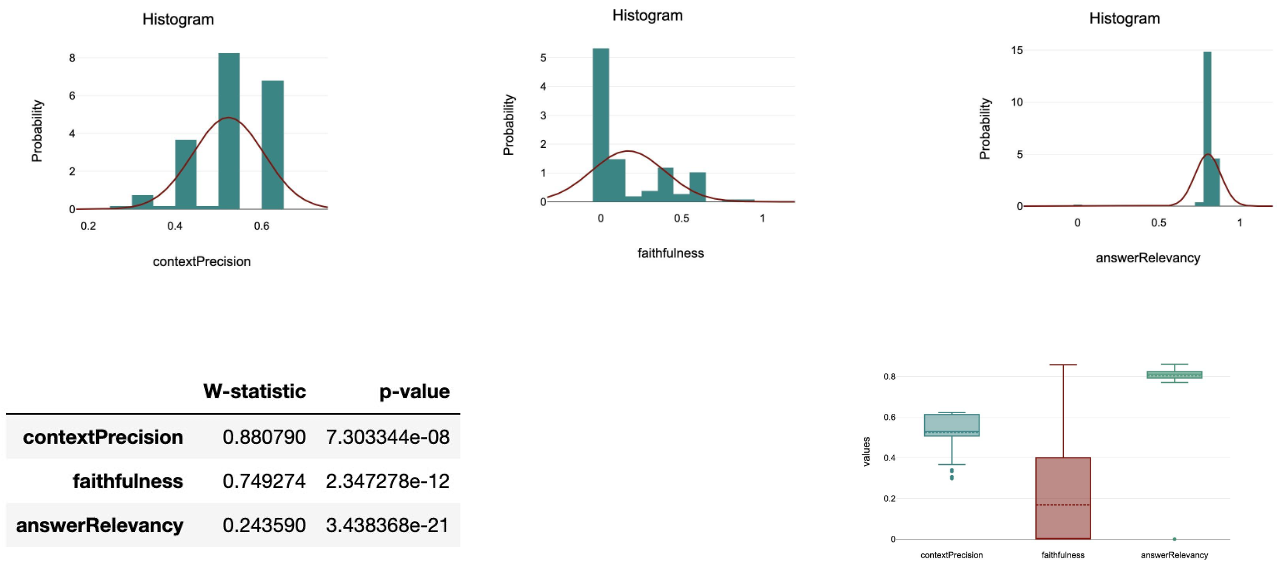}
\caption{This figure demonstrates the normality check for zero-shot rubric, all the LLM-evaluation metrics are not normally distributed.}
\label{nom2}
\centering
\end{figure}

In zero-shot rubric case \ref{nom2}, context precision histogram followed a roughly normal distribution, showing a slight right-skew. The faithfulness distribution is highly skewed toward 0, indicating that many responses had low factual alignment, with only a few higher values. The answer relevancy histogram is strongly right skewed, with a sharp peak near 1, suggesting that most responses were highly relevant. The p-values for Shapiro–Wilk test are all less than 0.05, therefore they are not normally distributed either.

\begin{figure}[h]
\centering
\includegraphics[width=\linewidth]{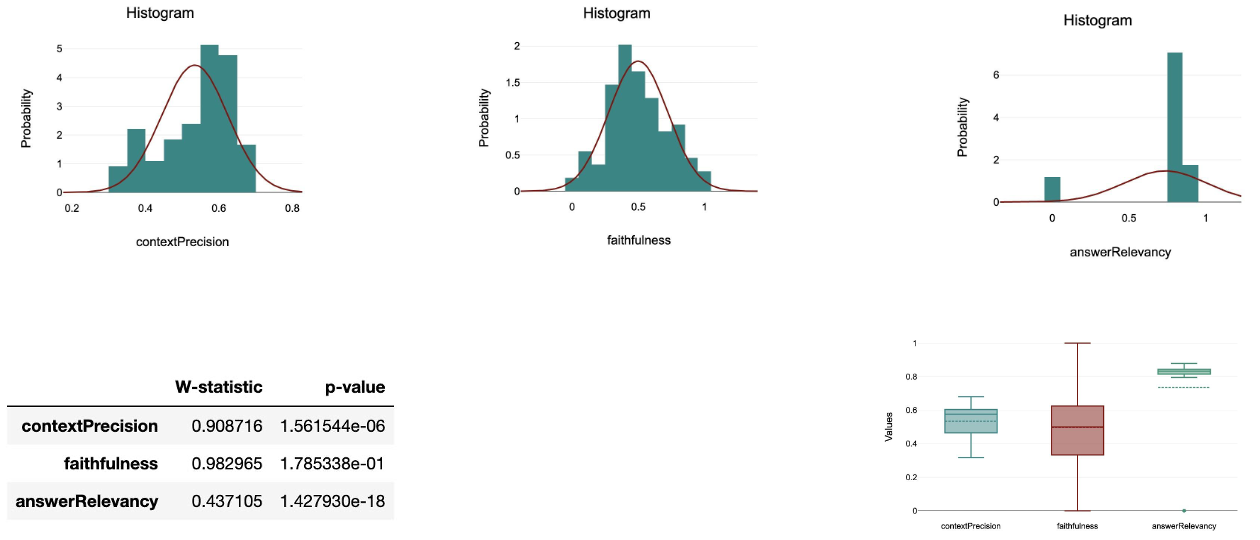}
\caption{This figure demonstrates the normality check for few-shot basic, only the faithfulness has a normal distribution, but the rest LLM-evaluation metrics are not normally distributed.}
\label{nom3}
\centering
\end{figure}

For the few shot basic case \ref{nom3}, context precision histogram followed a roughly normal distribution, suggesting that most values are concentrated in this range. The faithfulness distribution is right-skewed, indicating a tendency toward lower factual accuracy. The answer relevancy histogram is bimodal, showing that responses are either highly relevant or not relevant at all. The p-value for faithfulness is greater than 0.05, which means its distribution is normal, but the other two metrics are not since their p-value is less than 0.05.

Finally, it's the few-shot rubric case. Context precision histogram followed a roughly normal distribution, indicating relatively consistent precision scores. The faithfulness distribution is right-skewed, suggesting that many responses had low to moderate factual accuracy. The answer relevancy histogram is bimodal. The p-values for the Shapiro–Wilk test are all less than 0.05, which means their distributions are not normal.

\begin{figure}[h]
\centering
\includegraphics[width=\linewidth]{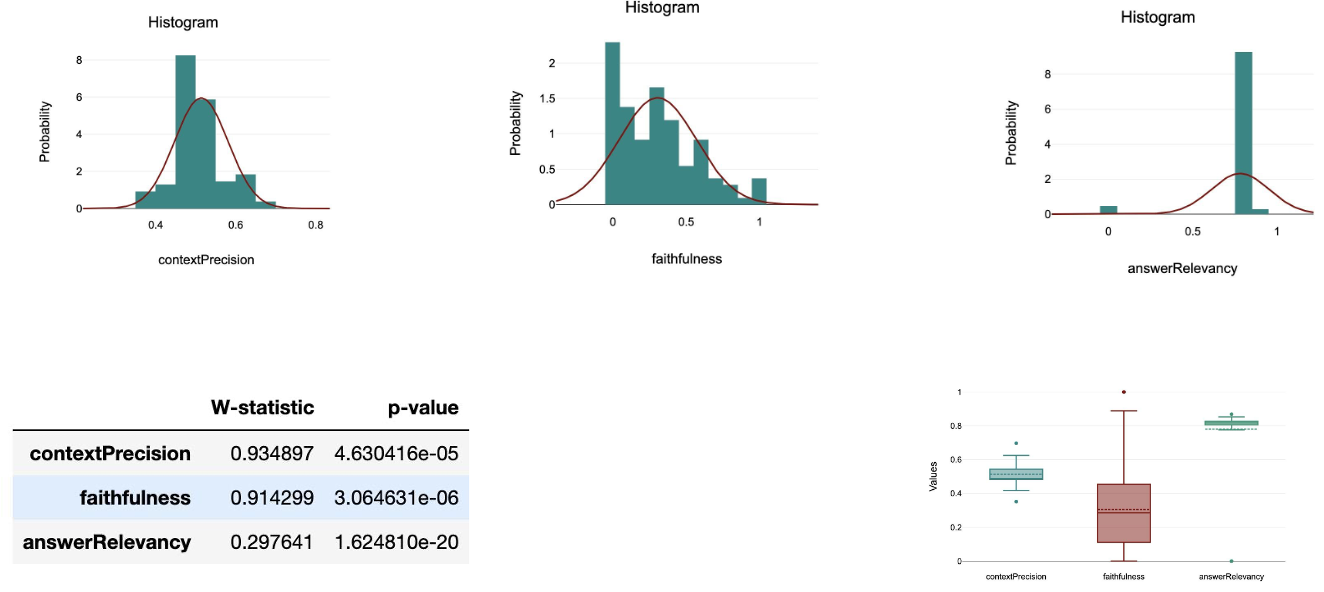}
\caption{This figure demonstrates the normality check for few-shot rubric, all the LLM-evaluation metrics are not normally distributed.}
\label{nom4}
\centering
\end{figure}

\paragraph{Nonparametric Test}
As shown in the previous section, since most of the distributions are not normal, we were not able to perform a parametric test to detect differences between each strategy. Therefore we performed the Kruskal-Wallis Test, which is non-parametric to compare non-normal distributed data by using rankings \cite{mckight2010kruskal}.

\begin{figure}[h]
\centering
\includegraphics[width=\linewidth]{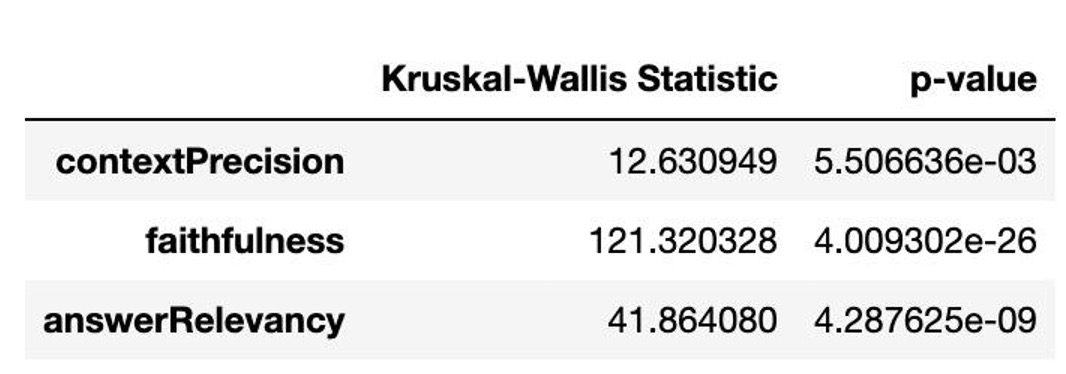}
\caption{This figure is the results of Kruskal-Wallis Test, since all the p-values are less than 0.05, it is likely to have difference between different prompting strategies for all three LLM evaluation metrics.}
\label{kw}
\centering
\end{figure}

As the result shown in \ref{kw}, the p-values for all three metrics were all below the 0.05 significance threshold, showing particularly strong significance. This indicates that the differences observed among the groups are unlikely to be due to random chance. The Kruskal-Wallis statistic values further suggest that Faithfulness exhibits the most substantial variation across groups, followed by answer relevancy and context precision.

\begin{figure}[h]
\centering
\includegraphics[width=\linewidth]{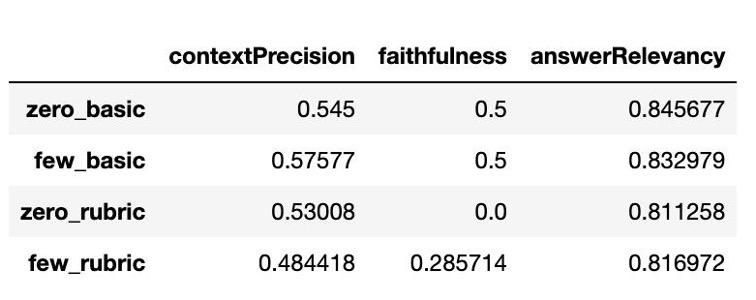}
\caption{This figure lists the medians for each LLM-evaluation metric by using each prompting strategy.}
\label{med}
\centering
\end{figure}

Furthermore, to conduct the post hoc analysis, we performed Dunn’s test for each LLM evaluation metric to determine which specific groups differed significantly from each other. Dunn’s test is a non-parametric pairwise comparison method that accounts for multiple comparisons while maintaining statistical rigor \cite{dinno2015nonparametric}. This test allowed us to identify which prompting strategies exhibited statistically significant differences in context precision, faithfulness, and answer relevancy. Afterward, a median analysis was conducted to determine which strategy yields higher values \cite{dinno2017package}, the values are shown in Figure \ref{med}.

\begin{figure}[h]
\centering
\includegraphics[width=\linewidth]{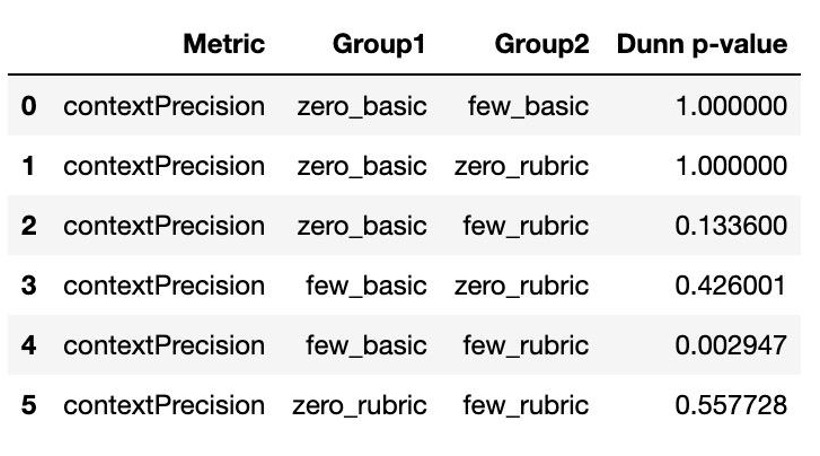}
\caption{This figure is the results of Dunn's Test for context precision, only one pair of comparison has significant differences.}
\label{context}
\centering
\end{figure}

Firstly, as shown in Figure \ref{context}, for context precision, the only pair of comparison that had a p-value less than significance level (0.05) is few-shot basic and few-shot rubric, suggesting that the inclusion of a rubric in a few-shot setting impacted context precision. Other comparisons had no significant difference between these groups. Based on the medians in Figure \ref{med}, few shot basic is more likely to yield higher context precision than few shot rubric.

\begin{figure}[h]
\centering
\includegraphics[width=\linewidth]{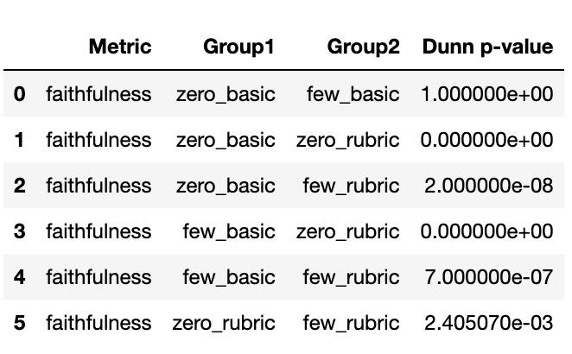}
\caption{This figure is the results of Dunn's Test for faithfulness, and there are five pairs of comparisons have significant differences. }
\label{faith}
\centering
\end{figure}

In Figure \ref{faith}, for faithfulness, there are five pairs of comparisons with significant differences, and the only pair with no significant difference is between zero-shot basic and few-shot basic. After analyzing the medians, zero-shot basic tended to have significantly higher faithfulness scores compared to zero-shot rubric and few-shot rubric, suggesting that adding a rubric in both settings may reduce faithfulness. Similarly, few-shot basic is significantly higher than both zero-shot rubric and few-shot rubric, indicating that providing more examples without a rubric improves faithfulness compared to rubric-based approaches. Additionally, zero-shot rubric tended to be significantly higher than few-shot rubric, further supporting that rubric-based prompting in a few-shot setting may negatively impact faithfulness.

\begin{figure}[H]
\includegraphics[width=15cm]{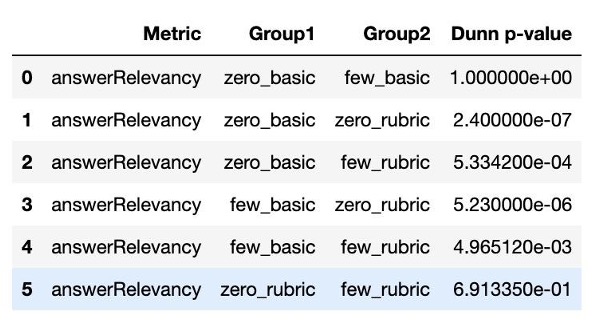}
\centering
\caption{This figure is the results of Dunn's Test for answer relevancy, and only two pairs have no significant differences.}
\label{anss}
\centering
\end{figure}

Finally, looking at answer relevancy in Figure \ref{anss}, the only two pairs with no significant differences are zero-shot basic vs. few-shot basic and zero-shot rubric vs. few-shot rubric, indicating that in both rubric and basic settings, adding examples did not impact answer relevancy. Otherwise, considering the medians, zero-shot basic has significantly higher answer relevancy than both zero-shot rubric and few-shot rubric, suggesting that introducing a rubric in both zero-shot and few-shot settings may reduce answer relevancy. Similarly, few-shot basic outperformed both zero-shot rubric and few-shot rubric. These findings suggested that the inclusion of a rubric may introduce constraints that negatively impact how directly the model addresses the question, whereas basic prompting strategies tended to yield more relevant responses.

% medians

\subsection{Summary}

\subsubsection{RQ1}
The analysis of distance correlation between sketch features from the synthetic dataset and LLM evaluation metrics suggested that sketch features have a generally weak influence on the interpretability of LLM-generated feedback. Across all prompting strategies, most sketch features exhibited negligible correlations with context precision, faithfulness, and answer relevancy.

\subsubsection{RQ2}

Through non-parametric tests, we can derive several insights regarding the trade-offs among different prompting strategies for LLMs to evaluate a synthetically labeled image dataset. Few-shot basic tended to achieve higher context precision than few-shot rubric, suggesting that adding a rubric may not enhance how well the LLM extracts relevant information. For Faithfulness, both zero-shot basic and few-shot basic perform better than zero-shot rubric and few-shot rubric, indicating that incorporating a rubric did not necessarily improve the factual accuracy of the generated feedback. Similarly, for answer relevancy, adding a rubric did not appear to enhance how well the LLM-generated response aligns with the input question. Overall, these findings suggest that introducing a rubric in prompting strategies may not improve LLM interpretability for evaluating sketch-based image labeling, and further research is needed to determine whether incorporating examples may improve LLM interpretability.

\section{Future Work}
\subsection{Limitations}

This research has some limitations. Firstly, since we used synthetic data for sketch-based labeling in images, it did not fully replicate human drawing behavior. One key issue is that the density of data points in the synthetic dataset does not align with natural human sketching tendencies, where drawing speed typically decreases in corners but increases during stroke rotations \cite{sezgin2007sketch}. Additionally, the dataset only accounts for right-handed drawers, limiting its applicability to a broader population. Another major limitation is that the synthetic dataset does not include temporal data, preventing analysis of motion-based features such as drawing speed and acceleration \cite{taranta2016rapid}. Lastly, the dataset itself had limited variety, because all the objects to label are all human, which may restrict the generalization of the findings and their applicability to diverse sketching styles and conditions.

\subsection{Future Directions}
In future research, it is important to enhance the synthetic dataset and incorporate real human data for a more comprehensive analysis. One key improvement is generating temporal information in synthetic data, which would enable researchers to extract motion-based features such as drawing speed and acceleration, providing deeper insights into sketch dynamics. Moreover, conducting a user study to collect real human-drawn sketches is essential for comparing the differences between synthetic and human data. This comparison would help identify discrepancies in drawing behavior, such as stroke density patterns and variability in movement. Also, analyzing human-generated sketches would offer valuable insights into natural drawing tendencies, enabling the refinement of synthetic datasets to better simulate real-world scenarios. Finally, through user studies, we will be able to obtain user's feedback about sketch-based interaction and attitudes toward the interpretability of LLM-generated text, which will further verify the validity and feasibility of  LLM-powered sketch-based image labeling system.

\section{Conclusion}
In this research, we explored the influence of SR features from a synthetic dataset on the interpretability of LLM-generated feedback and examined the trade-offs among different prompting strategies in evaluating images labeled by free-hand sketches. Through correlation analysis, the findings indicate that sketch features exhibit mostly weak or negligible correlations with LLM evaluation metrics including context precision, faithfulness, and answer relevancy. These results suggest that while certain sketch attributes, such as stroke density, length, and angular features, may have a minor influence on LLM-generated feedback, their overall impact remains limited.

In evaluating different prompting strategies, the study reveals that adding a rubric does not necessarily improve interpretability metrics. Specifically, few-shot basic outperforms few-shot rubric in context precision, while zero-shot basic and few-shot basic consistently achieve higher faithfulness and answer relevancy compared to rubric-based strategies. This suggests rubrics may impose constraints that reduce alignment and factual accuracy in LLM-generated responses, but the benefit of incorporating examples requires further research. 

Despite these insights, this research has some limitations, including the reliance on synthetic data, which does not fully replicate human sketching behavior, and the absence of temporal data to analyze motion-based features. In the future, it is necessary to enhance synthetic datasets with time-based data and motion-based features. Meanwhile, the next step is to conduct user studies to collect real users’ sketch labeling data and their attitudes toward sketch-based interaction and LLM-generated feedback. This approach will allow for a comparison between human-drawn sketches and synthetic samples, providing deeper insights and verifying the feasibility of an LLM-powered sketch-based image labeling assistant.

%%
%% The next two lines define the bibliography style to be used, and
%% the bibliography file.
\bibliographystyle{ACM-Reference-Format}
\bibliography{main}

%%
%% If your work has an appendix, this is the place to put it.
\appendix

\appendix

\section{Prompts for Different Strategies}
\subsection{Zero-shot Basic}

\begin{itemize}
\item \textbf{Role: User}: "Hello, I’m a user of an image labeling platform using a stylus and tablet to label images. 
I have already done some sketches in the image, but I need your help to improve my work. My work can be seen in the image. I want my labeling to be precise. Please give me some feedback. Thanks!"

\item \textbf{Role: System}: "Hi there! You are a friendly tutor and an evaluator, and your task is to evaluate the labeled image and detect some abnormalities based on inputs. You will be given 1 images as input, which is labeled by a person by using some sketch interface by using pen and stylus. The labeling sketches in the sketch are in color RED. A well-labeled object in the image should have the sketch completely enclose the object, there should be little gap between the boundary of the object and sketches, and it not out of the image boundary, and also the sketch should not overlay the object needed to be labeled. Here are some expectations for your output message: If you think the image labeling by the users is not bad, encourage the user and let them keep doing the good work. If you think it is not good, remind the user of their issue, give them suggestions, and let them keep grinding their techniques. Provide some feedback to the user so that they can improve their labeling. Please be patient and make sure your feedback is short(less than three sentences) and straightforward. Do not use emoji. Thanks!"
\end{itemize}

\subsection{Zero-shot Rubric}

\begin{itemize}
\item \textbf{Role: User}: "Hello, I am a user of an image labeling platform using a stylus and tablet to label images. I have already done some sketches in the image, but I need your help to improve my work. My work can be seen in the image. I want to achieve the standard in the rubric. Please give me some feedback. Thanks!"

\item \textbf{Role: System}: "Hi there! You are a friendly tutor and an evaluator, and your task is to evaluate the labeled image and detect some abnormalities based on inputs. You will be given 1 images as input, which is labeled by a person by using some sketch interface by using pen and stylus. The labeling sketches in the sketch are in color RED. A labeled object in the image should have the sketch completely enclose the object and should be as close to the boundary of the object as possible and not out of the image boundary, and also the sketch should not lie on the object that needs to be labeled.

Here are some evaluation criteria:
If the sketches are not out of the image boundary, the labeling sketches receive a 1 score.
If the sketches basically enclosed the object, the labeling sketches receive another 1 score.
If there is no large gap between the sketch and the object in the image, the labeling sketches receive another 1 score.
If the sketches do not overlay the object in an image, the labeling sketches receive another 1 score.
     
Here are some expectations for your output message:
If the score is greater than 2, encourage the user and let them keep doing the good work. If the score is less than 2, remind the user of their issue and give them suggestions. Provide some feedback to the user so that they can improve their labeling. Please be patient and pilot, and make sure your feedback is short(less than three sentences) and straightforward. Do not use emoji. Thanks!"
\end{itemize}

\subsection{Few-shot Basic}

\begin{itemize}
\item \textbf{Role: User}: "Hello, I am a user of an image labeling platform using a stylus and tablet to label images. I have already done some sketches in the first image, but I need your help to improve my work. My work can be seen in the first image. The second image is a perfect image labeling example, please reference it. I want to be as good as the perfect example. Please give me some feedback and suggestions. Thanks!"

\item \textbf{Role: System}: "Hi there! You are a friendly tutor and an evaluator, and your task is to evaluate the labeled image and detect some abnormalities based on inputs. You will be given 2 images as input, the first one is labeled by a person by using some sketch interface using a pen and stylus, and the second one is a perfect image labeling example, please reference this example in judging and provide suggestions. The labeling sketches in the sketch are in color RED. A labeled object in the image should have the sketch completely enclose the object, and should be as close to the boundary of the object as possible and not out of the image boundary, and also the sketch should not lie on the object needed to be labeled.

Here are some expectations for your output message:
Based on the evaluation criteria above, compare the user's labeling with the perfect example image. If the score is greater than 2, encourage the user and let them keep doing the good work. If the score is less than 2, remind the user of their issue and give them suggestions. Provide some feedback to the user and tell the user how they can improve their labeling and to be as good as the perfect example. Please be patient, and make sure your feedback is short(less than three sentences) and straightforward. Do not use emoji. Thanks."

\item \textbf{Referenced Example}: Figure \ref{ans}

\begin{figure}[H]
\includegraphics[width=10cm]{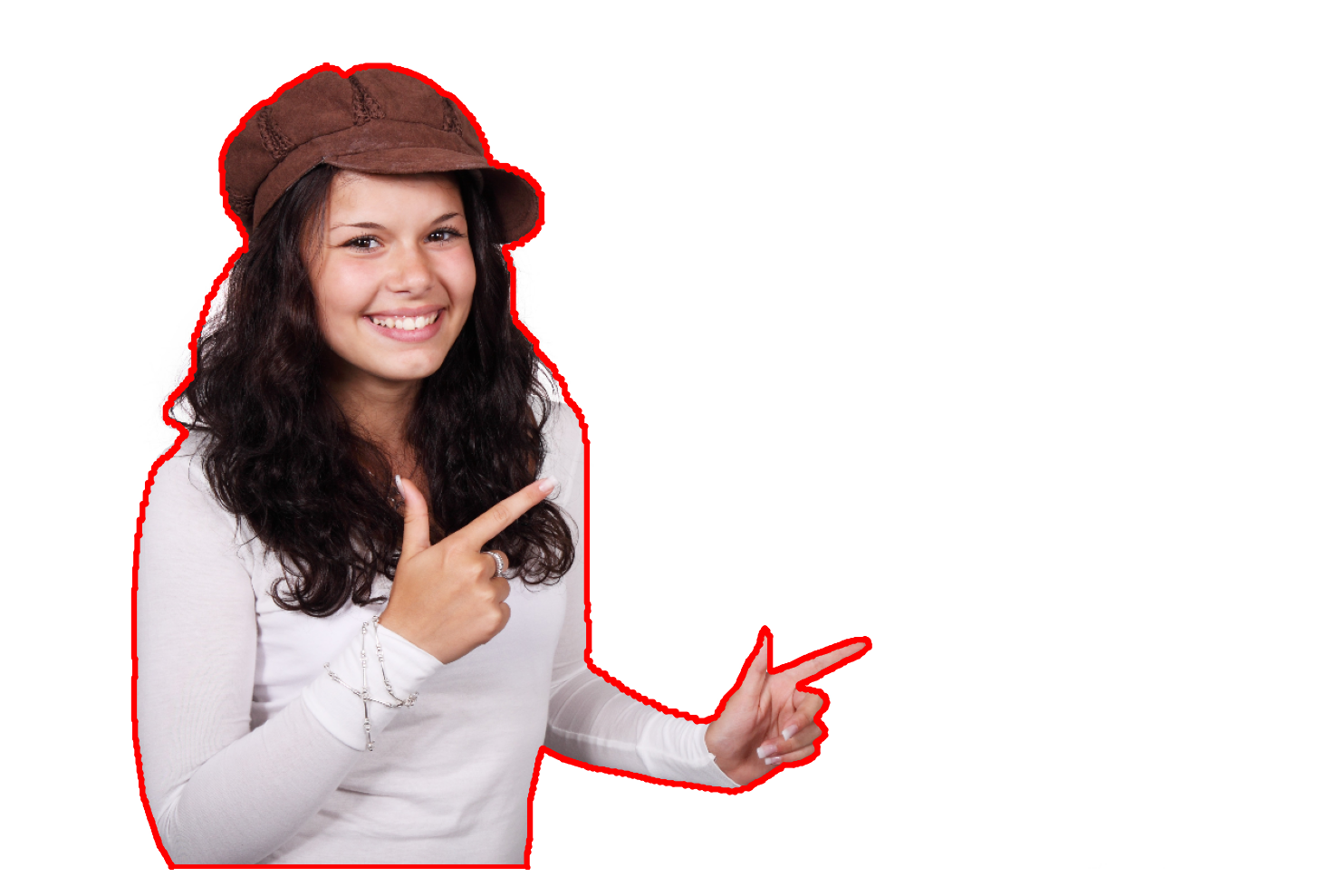}
\centering
\caption{This figure is a referenced example labeled image used in few-shot prompts.}
\label{ans}
\centering
\end{figure}

\end{itemize}

\subsection{Few-shot Rubric}

\begin{itemize}
\item \textbf{Role: User}: "Hello, I am a user of an image labeling platform using a stylus and tablet to label images. I have already done some sketches in the first image, but I need your help to improve my work. My work can be seen in the first image. The second image is a perfect image labeling example, please reference it. I want to achieve the standard in the rubric and to be as good as the perfect example. Please give me some feedback and suggestions. Thanks!"

\item \textbf{Role: System}: "Hi there! You are a friendly tutor and an evaluator, and your task is to evaluate the labeled image and detect some abnormalities based on inputs. You will be given 2 images as input, the first one is labeled by a person by using some sketch interface using a pen and stylus, and the second one is a perfect image labeling example, please reference this example in judging and provide suggestions. The labeling sketches in the sketch are in color RED. A labeled object in the image should have the sketch completely enclose the object, and should be as close to the boundary of an object as possible and not out of the image boundary, and also the sketch should not lie on the object needed to be labeled.

Here are some evaluation criteria:
If the sketches are not out of the image boundary, the labeling sketches receive a 1 score.
If the sketches basically enclosed the object, the labeling sketches receive another 1 score.
If there is no large gap between the sketch and the object in the image, the labeling sketches receive another 1 score.
If the sketches do not overlay the object in an image, the labeling sketches receive another 1 score.
   
Here are some expectations for your output message:
Based on the evaluation criteria above, compare the user's labeling with the perfect example image.
If the score is greater than 2, encourage the user and let them keep doing the good work. If the score is less than 2,
remind the user of their issue and give them suggestions.
Provide some feedback to the user and tell the user how they can improve their labeling and be as good as the perfect example. Please be patient,and make sure your feedback is short(less than three sentences) and straightforward. Do not use emoji. Thanks!"

\item \textbf{Referenced Example}: Figure \ref{ans}
% \begin{figure}[H]
% \includegraphics[width=10cm]{figures/pef.png}
% \centering
% \caption{Referenced Example Labeled Image}
% \label{ans}
% \centering
% \end{figure}
\end{itemize}

\end{document}